\documentclass[a4paper,12pt,final]{article}

\usepackage[sort&compress]{natbib}
\setcitestyle{open={},close={},super}
\usepackage{hyperref}
\usepackage{amsmath}
\usepackage[T1]{fontenc}       
\usepackage{graphicx}
\usepackage{authblk}
\usepackage{bookmark}
\usepackage{bm}
\usepackage{upgreek} 
\usepackage[utf8]{inputenc} 
\usepackage{textcomp}

\author[1]{Armin Feist\footnote{armin.feist@uni-goettingen.de}}
\author[1]{Sergey V. Yalunin}
\author[2]{Sascha Sch\"afer}
\author[1]{Claus Ropers\footnote{claus.ropers@uni-goettingen.de}}

\affil[1]{University of G\"ottingen, IV. Physical Institute, G\"ottingen, Germany}
\affil[2]{Carl von Ossietzky Universit\"at Oldenburg, Institute of Physics, Oldenburg, Germany}

\date{}

\title{High-purity free-electron momentum states prepared by three-dimensional optical phase modulation}

\begin{document}
	
\maketitle

\addcontentsline{toc}{section}{Abstract}
\begin{abstract}
We demonstrate the quantized transfer of photon energy and transverse momentum to a high-coherence electron beam. In an ultrafast transmission electron microscope, a three-dimensional phase modulation of the electron wavefunction is induced by transmitting the beam through a laser-illuminated thin graphite sheet. This all-optical free-electron phase space control results in high-purity superpositions of linear momentum states, providing an elementary component for optically programmable electron phase plates and beam splitters.
\end{abstract}


The coherent manipulation of particle wave functions forms the basis of nanoscale and interference-enhanced precision measurements, ranging from matter-wave interferometry \cite{Cronin2009, Hasselbach2010} and free-electron lasers \cite{Pellegrini2016}, to advanced electron imaging and holography \cite{Midgley2009}. In electron microscopy, phase plates and diffractive elements produce tailored quantum states, such as electron vortex beams \cite{Uchida2010, Verbeeck2010, McMorran2011, Grillo2014, Bliokh2017a} and self-healing Airy \cite{Voloch-Bloch2013} or non-diffractive Bessel beams \cite{Courvoisier2012, Kaminer2012, Grillo2014a}, with prospects for nanoscale chiral sensing \cite{Asenjo-Garcia2014,Lloyd2017} and manipulations \cite{Verbeeck2013, Verbeeck2018}.

Currently, optical field control of electron beams is experiencing growing interest from an applied and fundamental perspective \cite{GarciadeAbajo2010b, Hemsing2014, England2014, Pellegrini2016, Ciappina2017}. Recently, a Zernicke phase plate based on the ponderomotive potential in optical fields was demonstrated \cite{Muller2010, Schwartz2019}. Furthermore, high-intensity laser pulses in free space or enhanced by cavities, waveguides and nanostructures facilitate electron pulse characterization \cite{Bucksbaum1987, Siwick2005, Hebeisen2008, Gao2012, Kozak2017}, compression \cite{Kassier2012, Wong2015}, acceleration \cite{Breuer2013, Peralta2013}, and attosecond structuring \cite{Priebe2017, Morimoto2018, Kozak2017}. Electron phase-space manipulation within the cycle of the electromagnetic field is accessible in the Terahertz \cite{Wimmer2014, Nanni2015, Kealhofer2016, Curry2018, Zhang2018, Li2019, Zhang2019a, Kim2019, Zhao2019} and radiofrequency domains \cite{Chatelain2012, Gliserin2015, Maxson2017, Verhoeven2018, Jing2019}.

Ultrafast transmission electron microscopy (UTEM) \cite{Zewail2006, Flannigan2012, Feist2017}, an emerging approach to study nanoscale dynamics, is particularly suited for the study of coherent electron-light interactions, including near-field mediated inelastic scattering \cite{Barwick2009, Park2010, GarciadeAbajo2010,  Feist2015}. In UTEM, a number of experimental studies employed longitudinal momentum transfer to conduct near-field imaging \cite{Barwick2009, Piazza2015, Yurtsever2012, Feist2015, Piazza2015, Madan2019}, characterize ultrashort electron pulses \cite{Barwick2009, Kirchner2014, Feist2017}, carry out plasmon excitation spectroscopy \cite{Pomarico2018} and probe cavity modes \cite{Kfir2019, Wang2019a}. In these experiments, electron spectroscopy reveals quantized energy transfer in multiples of the photon energy. This direct access to the longitudinal phase space enabled the observation of free-electron Rabi oscillations \cite{Feist2015}, coherent quantum state control \cite{Echternkamp2016, Priebe2017, Vanacore2018} and attosecond bunching \cite{Priebe2017}.

Addressing the transverse beam direction, recent experiments reported on the optical deflection \cite{Vanacore2018, Krehl2018} and streaking \cite{Kozak2017, Morimoto2018}, and on vortex phase shaping \cite{Vanacore2019} by chiral plasmonic fields \cite{Kim2010, Spektor2017}. However, the observation of quantized momentum transfer induced by visible light is challenging. To date, experiments on the Kapitza-Dirac effect \cite{Freimund2001} remain the only demonstrations of a quantized transverse electron beam deflection by optical fields which leads to individual, angularly separated beams.

Here, we report the optical preparation and observation of high-purity quantum coherent linear momentum superposition states in the transverse and longitudinal direction of an electron microscope beam. Low-emittance ultrashort electron pulses, generated by nanotip photoemission, are collimated to a micrometer-scale transverse coherence length and transmitted through a laser-illuminated graphite thin film. The imprinted three-dimensional sinusoidal phase modulation yields a coherent superposition of correlated energy-momentum ladder states, which is mapped by its far-field scattering distribution.

\begin{figure*}
\includegraphics{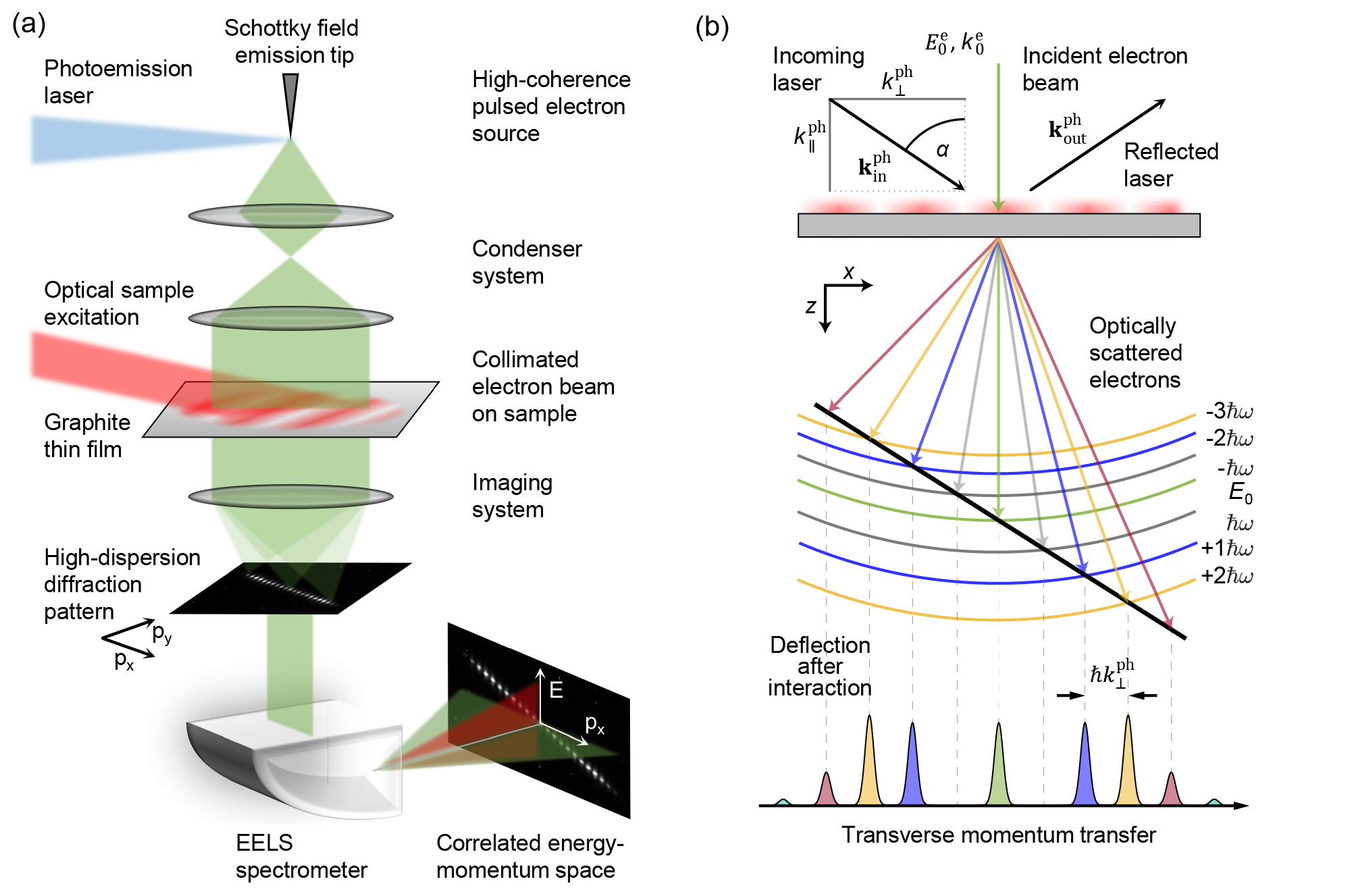}
\caption{\label{fig:f1} Schematic setup and concept of transverse momentum resolved inelastic electron-light scattering. (a) Experimental Setup. Inside of an ultrafast transmission electron microscope, photo-emitted ultrashort electron pulses are collimated and transmitted through a laser side-illuminated graphite thin film. After optical interaction, the electron beam is analyzed in terms of its distributions in reciprocal (transverse) and in energy (longitudinal) space. (b) Ewald-sphere construction of inelastic electron scattering at the laser-generated traveling phase grating. The induced three-dimensional optical phase modulation yields a high-purity momentum superposition state, with separations given by integer numbers of the transverse photon momentum $\hbar k_\perp^{\mathrm{ph}}$ and photon energy $\hbar\omega$ (centered around the initial electron kinetic energy $E_0^e$).}
\end{figure*}

We begin by considering the theoretical basis for our study. Inelastic electron-light scattering is well-understood, and several approaches have been presented to derive analytical expressions for PINEM sideband amplitudes, including a Green’s function formalism \cite{GarciadeAbajo2010}, direct integration of the Schrödinger equation \cite{Park2010} and a ladder operator formalism \cite{Feist2015}.

In the following, setting a general framework for transverse beam shaping, we offer an alternative, rather concise derivation using a path integral representation of the wavefunction \cite{Schulman2005}. In the semiclassical limit, it is expressed in terms of the classical action as $\Psi(\pmb{r},t) \sim e^{iS(\pmb{r},t)/\hbar}$, where $S(\pmb{r},t)=\int_{t_0}^t L[\pmb{r}(\tau)] d\tau$ is the action - a solution of the Hamilton-Jacobi equation, and $L[\pmb{r}(\tau)]$ is the Lagrangian. The integral over $\tau$ is evaluated along a specific classical path $\pmb{r}(\tau)$ that ends at the point $\pmb{r}(t)=\pmb{r}$. The initial point of the path at time $t_0$ is specified by the initial velocity $\dot{\pmb{r}}(t_0) = \pmb{v} = c^2\pmb{p}/E(\pmb{p})$. Consequently, the phase difference $\Phi(\pmb{r},t)$ between the phases of the initial plane-wave state and the final state at position $\pmb{r}$ and at time $t$ can be written as:
\begin{equation} \label{phase}
\Phi(\pmb{r},t)=\frac{1}{\hbar} \int_{t_0}^t \left\{L[\pmb{r}(\tau)] - \pmb{p}\cdot \dot{\pmb r}(\tau)-E(\pmb{p})\right\} d\tau,
\end{equation}
where the second and third terms subtract the phase acquired by the initial state along the path of integration. Equation~(\ref{phase}) is the exact semiclassical limit for the phase modulation. It can be further simplified under the assumption that the energy change is small compared to the initial kinetic energy of the electron. Treating the interaction as a perturbation, one obtains in the first order the following expression \cite{Lubk2018}:
\begin{equation}\label{phase2}
\Phi(\pmb{r},t)= -\frac{1}{\hbar} \int_{t_0}^t H_I(\pmb{r}(\tau),\tau) d\tau,
\end{equation}
where $H_I(\pmb{r},t)=\frac{ep}{m} A_z(\pmb{r},t)$ is the interaction Hamiltonian (with elementary charge $e$ and electron rest mass $m$), and $A_z(\pmb{r},t)$ is the vector potential in the propagation direction of the electron beam  (which we choose along $z$). With the same accuracy, $\pmb{r}(\tau)$ can be approximated by a straight-line path: $\pmb{r}(\tau) = \pmb{r}+\pmb{v}(\tau-t)$, where $\pmb{v}$ is the initial velocity. 

The spatial and temporal structure of the phase modulation depends on the specific field geometry in a given experiment. A field distribution breaking translational invariance along the electron propagation direction, e.g., at an optical nanostructure, can induce transitions between states separated by the photon energy \cite{Barwick2009}. For translational invariance in the direction perpendicular to the beam path, such transitions will entail quantized transverse momentum transfer \cite{GarciadeAbajo2016}. 
The simplest geometry fulfilling both requirements is given by light reflection from a planar surface \cite{Kirchner2014, Priebe2017, Vanacore2018, Morimoto2018}. This allows for a time-harmonic electromagnetic field that abruptly varies in the $z$-direction but is propagating along the $x$-axis.

Consequently, one can choose the form $A_z(\pmb{r},t) = G(z)/\omega \cdot \sin (\omega t- k^\mathrm{ph}_{\perp}x)$, where $G(z)$ denotes the $z$-dependent amplitude of the electric field, with angular frequency  and transverse wavevector component $k^\mathrm{ph}_{\perp}$.

Hereby, equation (2) can be concisely written in terms of the Fourier transform of $G(z)$, as follows:
\begin{eqnarray}\label{eqF2}
\Phi(\pmb{r},t) = \frac{ep}{m\hbar\omega} \textrm{Im} \int_{t_0}^t G(z+v[\tau-t])e^{i(k^\mathrm{ph}_{\perp}x-\omega \tau)} d\tau 
\nonumber\\
=2 |g| \sin(k^\mathrm{ph}_{\perp}x+\omega z/v -\omega t +\arg g),
\end{eqnarray}
where $g$ describes the strength of electron-light interaction:
\begin{equation}\label{g_equ}
g = \frac{e}{2\hbar\omega} \int_{-\infty}^{\infty} G(z) e^{-i\omega z/v} dz.
\end{equation}
It is convenient to deal with the phase modulation in a shifted frame (given by the substitution $z-vt \to z$ into Eq.~(\ref{eqF2})) because, in this frame, both the phase and the initial state $\Psi_0(\pmb{r},t)=\exp[i(pz-Et)/\hbar]$ are independent of time $t$. Finally, we arrive at
\begin{equation}\label{phase3}
\Phi(\pmb{r}) = 2 |g|  \sin(k^\mathrm{ph}_{\perp}x+\omega z/v+\arg g)
\end{equation}
Electron scattering at such a time-dependent phase grating results in the formation of distinct photon orders $N$ in far-field diffraction, with amplitudes given by Bessel functions of the first kind $P_N = J_{N}^{2}(2|g|)$ \cite{Park2010, Feist2015}. Physically, the absorption of a photon with energy $\hbar\omega$ necessarily involves also absorption of its transverse momentum $p^\mathrm{ph}_{\perp}=\hbar k^\mathrm{ph}_{\perp}=h/\lambda\cdot\sin\alpha$, with wavelength $\lambda$ and angle of incidence $\alpha$ for a sample plane normal to the electron beam (cf. Fig.~\ref{fig:f1}b). As a consequence, a one-to-one correspondence of the scattering distribution in the total kinetic energy and the transverse momentum subspace is expected (Fig.~\ref{fig:f1}b) \cite{GarciadeAbajo2016}. We note that transverse optical phase modulation with a uniform coupling constant $g$ acts as a coherent inelastic beam splitter for free-electron beams, if the individual diffraction orders are fully separated in angle. For visible optical frequencies and near-relativistic electrons, the corresponding angular separation amounts to only a few microradians. Therefore, the experimental characterization of such phase-shaped beams, and in particular the resolution of individual scattering orders, requires electron beams of sufficiently low divergence, connected to a micrometer-scale transverse coherence length $\xi_{c}=\lambda_{e}/(2\pi\cdot\sigma_\theta)$, with root-mean-square (rms) angular spread $\sigma_\theta$ \cite{VanOudheusden2007}.

As we now demonstrate, such conditions are achieved in our ultrafast transmission electron microscope, which is based on laser-driven photoelectron emission from a Schottky field emission source \cite{Feist2017, Bach2019}.

\begin{figure*}
\begin{center}
\includegraphics{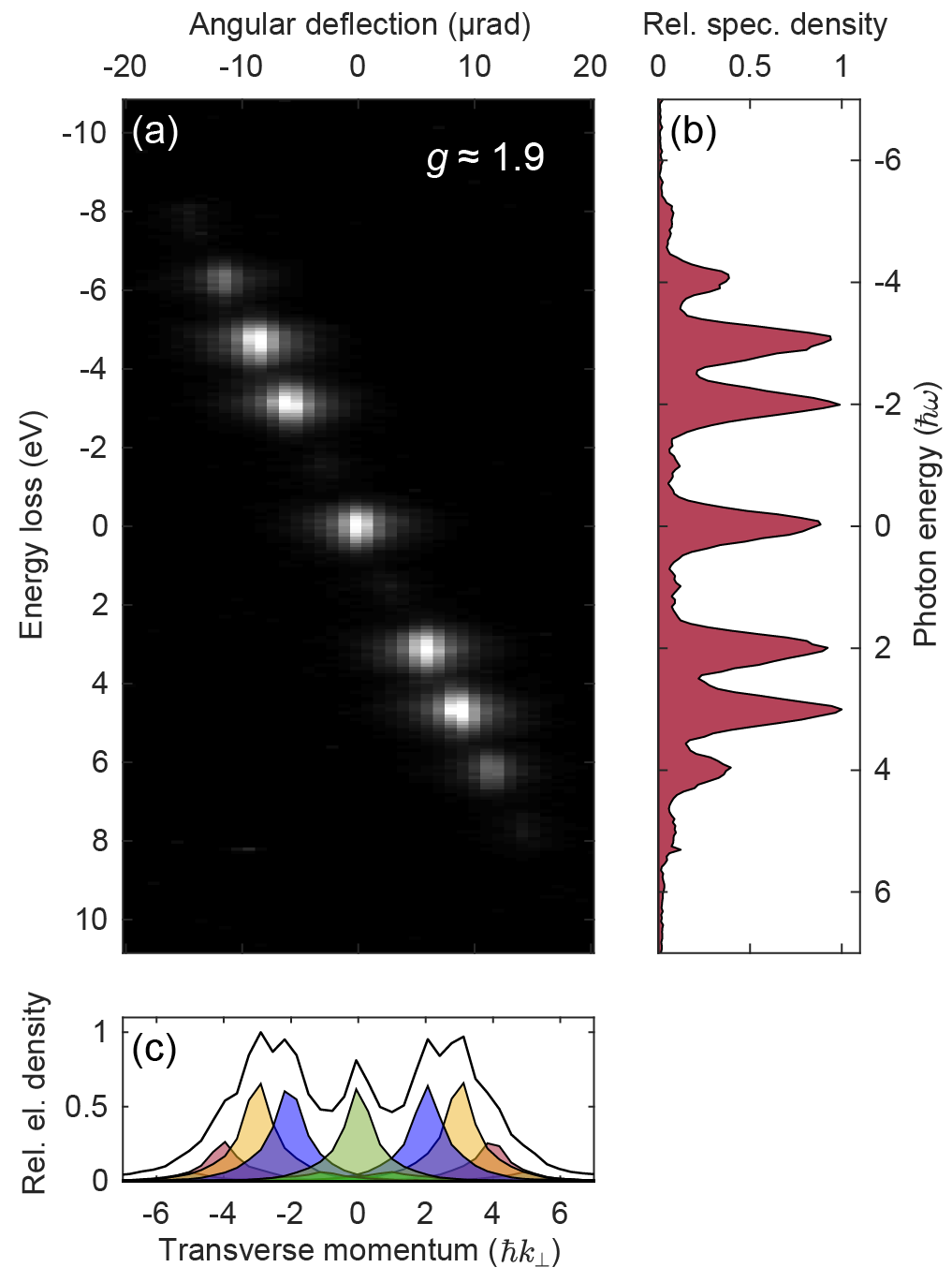}%
\caption{\label{fig:f2} Experimental energy-momentum correlation map. a) Far-field electron scattering distribution resolved in energy and transverse momentum space (oriented along the optical plane of incidence) with normalized orthogonal projections b) an c) (coupling constant $g \approx 1.9$). b) Spectral density as a function of the photon energy $\hbar\omega$. c) Scattering distribution in transverse phase space with angular resolution limited by the employed electron energy loss spectrometer. The contributions of individual states are color-coded.}
\end{center}
\end{figure*}

\begin{figure*}
\includegraphics{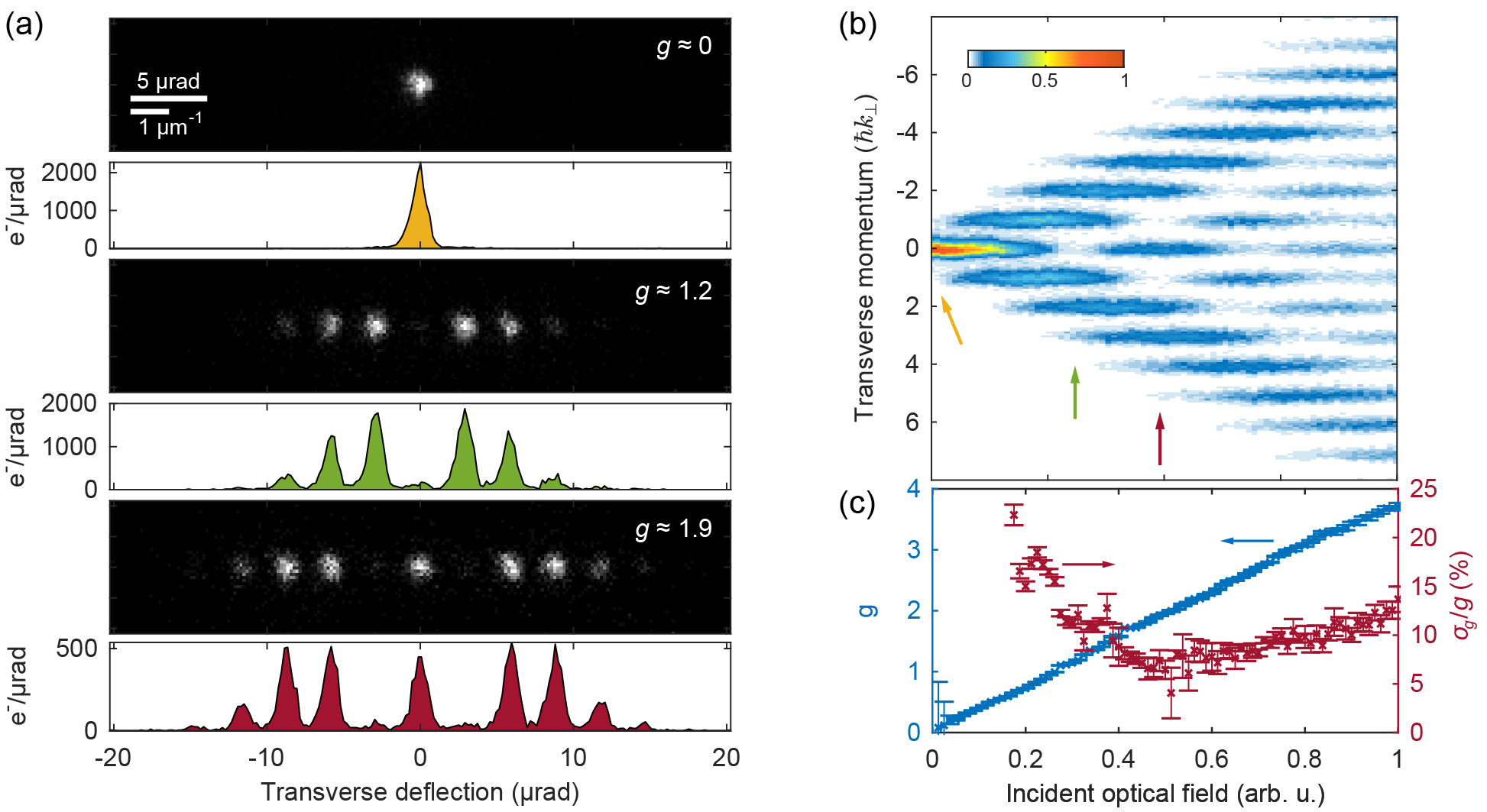}
\caption{\label{fig:f3} Optical power-dependent far-field scattering distribution. a) Single-electron counted high-dispersion diffraction patterns for varying optical field strength (coupling constant $g$ indicated in individual subpanels) and corresponding projection into the $k_x$-subspace. b) Normalized population of transverse momentum states as a function of optical field strength. The diffraction pattern shown in a) are indicated by arrows. c) Extracted electron-light coupling strength, assuming a Gaussian distribution with mean $g$ (blue) and standard deviation $\sigma_{g}$ (red).}
\end{figure*}

In the experiment (Fig.~\ref{fig:f1}a), we generate ultrashort, high-coherence electron pulses by localized single-photon photoemission from a nanoscopic, Schottky-type field emitter tip \cite{Feist2017}. After acceleration to a kinetic energy of \mbox{200 keV} (2.51-pm de-Broglie wavelength), the electron beam is collimated to a FWHM full beam divergence of \mbox{$\Delta{}\theta=1.0$ µ$\mathrm{rad}$}, resulting in a lateral coherence length of \mbox{$\xi_{c}=0.93$ µm} (cf. Fig.~\ref{fig:f3}a, top), and radially limited to a diameter of about \mbox{13 µm}. Passing the electrons through the optical field of ultrashort laser pulses (\mbox{800-nm} center wavelength, \mbox{$\sim{}50$ µm} focal spot size, up to \mbox{150-nJ} pulse energy, dispersively stretched to 3.4-ps duration), that are reflected from the surface of a 120-nm thick single crystalline graphite flake (\mbox{882.5-nm} effective period of the transverse optical phase grating) \footnote{The sample is prepared by cleaving a graphite single crystal perpendicular to the c-axis. In the experiment, the relative angle between laser and electron beam is \mbox{$\alpha=59^\circ$} and the sample was tilted away from the surface normal and laser incidence by \mbox{$\alpha=3.5^\circ$}, precisely determined by convergent beam electron diffraction}, imprints a sinusoidal phase modulation onto the electron wavefunction in the transverse and longitudinal beam directions (Fig.~\ref{fig:f1}b).

The scattered electron distribution is analyzed by operating the electron microscope in diffraction mode (effective camera length of \mbox{33 m}). We record maps of kinetic energy and momentum using an electron spectrometer (Fig.~\ref{fig:f2}), or of the full transverse momentum distribution in the far-field (Fig.~\ref{fig:f3}).

Figure ~\ref{fig:f2}a shows the kinetic energy and angle-resolved far-field scattering distribution \footnote{diffraction angles calibrated with a grating replica of \mbox{463-nm} spacing}, clearly demonstrating the creation of an equally-spaced comb of energy-momentum states (Fig.~\ref{fig:f2}b). The transitions in this two-dimensional energy-momentum ladder (cf. Fig.~\ref{fig:f1}b) lead to a correlated quantized gain or loss of kinetic energy $\hbar\omega$ and transverse photon momentum $p^\mathrm{ph}_{\perp}=\hbar k^\mathrm{ph}_{\perp}$ (Fig.~\ref{fig:f2}c), with a corresponding single-order deflection angle of \mbox{$\theta_{\perp} =\arctan(p^\mathrm{ph}_{\perp}/p_0^\mathrm{e}) = 2.90$ µrad}.

Notably, while the sidebands are fully separated in energy, with a visibility of\ {$\hbar\omega/\Delta E= 3.1$} ($\Delta E$ - FWHM energy spread of the electron beam), the optics of the EEL spectrometer and the employed scintillator-coupled CCD camera limit the detected angular separation to \mbox{$k^\mathrm{ph}_{\perp} / \Delta{}k^\mathrm{EELS}_{\perp} = 0.9$} ($\Delta{}k^\mathrm{EELS}_{\perp}$ - FWHM of detected transverse electron momentum distribution).

Facilitated by the low-divergence electron beams used, in the following, we directly resolve individual scattering orders by their deflection angles, and without employing energy-dispersive electron optics. We map the full two-dimensional transverse phase space by recording diffraction patterns after the optical interaction. Figure ~\ref{fig:f3} shows the far-field scattering distribution using the same experimental conditions as for Fig.~\ref{fig:f2}, but recording the diffracted beam by counted electron imaging with a direct detection CMOS camera (Direct Electron DE16). For increasing optical fields (cf. Fig.~\ref{fig:f3}a), the initial single-peak transverse momentum distribution splits into an equally spaced comb, separated by the transverse photon momentum \mbox{$p^\mathrm{ph}_{\perp}=\hbar k^\mathrm{ph}_{\perp}$} and well resolved with a visibility of $k^\mathrm{ph}_{\perp} / \Delta{}k^\mathrm{e}_{\perp} = 2.9$, with ($\Delta{}k^\mathrm{e}_{\perp}$ - FWHM transverse momentum spread of the electron beam). 
The purity of the quantum superposition state prepared at a single optical interaction is encoded in the optical-field dependent sideband populations (Fig.~\ref{fig:f3}b, top), which are, for a non-mixed state, analytically given by Bessel functions of the first kind $J_N(2g)$ \cite{Park2010, Feist2015}. The experimentally observed occupations are well reproduced by considering a Gaussian distributed coupling constant $g$ with standard deviation $\sigma_g(g)$ (Fig.~\ref{fig:f3}c). Beyond a value of $g=2$, we find a relative uncertainty $\sigma_g/g$ between $6\%$ and $12\%$, which indicates a high purity of the generated superposition quantum state \footnote{Lower $g$ values and a smaller number of populated states lead to a higher relative uncertainty in the fit}. A unique fingerprint of such a state -- and validation of its action as a coherent electron beam splitter -- is the full suppression of specific sideband orders, e.g., in the zero order beam ($N=0$) for the first zero crossing of $J_{0}(2g)$, which is achieved for $g\approx1.2$, as clearly visible in Fig.~\ref{fig:f3}a (middle).

Generally, multiple sources of incoherence are of relevance, which reduce the purity of the quantum superposition state. In the current experiment, there are two main contributions, which are minimized by the design of the experiment. Firstly, incoherent ensemble averages in the preparation of the initial quantum state (i.e., the initial spread in transverse momentum and energy), are reduced by collimating our high-coherence beam, enabling a full separation of the photon orders. Secondly, temporal and spatial variations in the field strength lead to averaging in $g$, which we minimize by limiting the electron beam diameter and temporally stretching the optical pulse. The simultaneous need of a low electron beam divergence and small diameter require a picometer-scale transverse beam emittance $\varepsilon_{x,y}$. In our experiment, we achieve a value of $\varepsilon_{x,y}=\beta\gamma\cdot\sigma_{x}\sigma_{\theta}=1.63~\mathrm{pm\cdot rad}$ (apertured, at the sample stage, with relativistic parameters $\beta$ and $\gamma$). Such high-brightness/low-emittance ultrashort electron pulses are readily available in field emitter-based UTEM instruments \cite{Feist2017, Houdellier2018, Verhoeven2018, Zhu2020}.

In conclusion, we have realized a coherent inelastic beam splitter for free-electron beams and demonstrate the optical preparation of high-purity transverse electron beam states.

These capabilities extend the toolset of matter-wave interferometry by providing coherent optical beam manipulation, e.g., as an element for new forms of interference-enhanced multi-path electron microscopes \cite{Kruit2016}. Furthermore, taking advantage of the spectral state separation, inelastic electron holography in a STEM-type geometry \cite{Yasin2018a} may give access to non-local information in fundamental material excitations \cite{Lichte2000, GarciadeAbajo2010}. Exploiting the advanced transverse phase-space control in the sub-optical cycle temporal structuring of electron beams will facilitate electron microscopy with isolated attosecond electron pulses, reminiscent of the attosecond lighthouse in optics \cite{Kim2013}. Finally, the concept of high-coherence transverse and longitudinal phase modulation can be extended to complex plasmonic light fields, thus, providing for an externally programmable three-dimensional all-optical phase plate for electrons.


\addcontentsline{toc}{section}{Acknowledgement}
\section*{Acknowledgement}
We thank the Göttingen UTEM team for ongoing support and fruitful collaboration. We acknowledge insightful discussions with Katharina\,E. Priebe, Thomas Rittmann, Nara Rubiano da Silva, Tyler Harvey and Hugo Lourenço-Martins. This work was funded by the Deutsche Forschungsgemeinschaft (DFG) in the Collaborative Research Center “Atomic Scale Control of Energy Conversion” (DFG-SFB 1073, project A05), the Priority Program “Quantum Dynamics in Tailored Intense Fields” (DFG-SPP 1840) and via funds of the Leibniz Program.


\bibliographystyle{ieeetr}
\bibliography{literature_arXiv}

\end{document}